\documentclass[finallayout,an,fleqn]{w-art}
\usepackage{times}
\usepackage{w-thm}
\usepackage{isolatin1}
\theoremstyle{plain}

\theoremstyle{definition}

\usepackage[]{graphicx}
\usepackage{isolatin1}
\chardef\bslash=`\\ 

\begin{document}
\DOIsuffix{theDOIsuffix}
\Volume{324}
\Issue{S1}
\Copyrightissue{S1}
\Month{01}
\Year{2003}
\pagespan{3}{}
\Receiveddate{15 November 2002}
\Reviseddate{30 November 2002}
\Accepteddate{2 December 2002}
\Dateposted{3 December 2002}



\title[Warm molecular gas, dust and ionized gas in the
500 central pc of the Galaxy]
{Warm molecular gas, dust and ionized gas in the 
500 central pc of the Galaxy}


\author[Rodríguez-Fernández]{N.J. Rodríguez-Fernández
   \footnote{e-mail: {\sf nemesio.rodriguez@obspm.fr},
             Phone: +33\,140\,512\,061, Fax: +33\,140\,512\,002}\inst{1}}
        \address[\inst{1}]{LERMA, Observatoire de Paris,
	61, Av de l'Observatoire, 75014 Paris, France}
\author[Martín-Pintado]{J. Martín-Pintado\inst{2}}
\address[\inst{2}]{Instituto de Estructura de la Materia, CSIC, 
          Serrano 121, 28006, Madrid, Spain }
\author[Fuente]{A. Fuente\inst{3}}
\address[\inst{3}]{Observatorio Astronómico Nacional, IGN, Apdo. 1143,
28800 Alcalá de Henares, Spain}
\author[Wilson]{T.L. Wilson\inst{4}}
\address[\inst{3}]{Max-Planck-Institut für Radioastronomie,
Auf dem H\"ugel 69, 53121 Bonn, Germany}               
\begin{abstract}
We present infrared and millimeter observations  of
molecular gas, dust and ionized gas towards a sample of clouds
distributed along the 500 central pc of the Galaxy.
The clouds were selected to investigate the physical state, in 
particular the high gas temperatures, of the Galactic center region (GCr)
clouds located far from far-infrared of thermal radio continuum sources.
We have found that  there is ionized gas  associated with the molecular gas.
The ionizing radiation is hard ($\sim 35000$ K) but diluted due
to the inhomogeneity of the medium.
We estimate that $\sim 30 \%$ of the warm molecular
gas observed in the GCr clouds
is heated by ultra-violet radiation in photo-dissociation regions.
\end{abstract}
\maketitle                   




\renewcommand{\leftmark}
{Rodríguez-Fernández et al.: Warm molecular gas, dust and ionized gas in the
500 central pc of the Galaxy}


\section{Introduction}
\label{required}
The interstellar medium in the 500 central pc of the Galaxy (hereafter
Galactic Center region, GCr) is mainly molecular.
The molecular clouds in the GCr exhibit an extended gas component with
high temperature (150 K).
On the contrary, the dust temperature is lower than $\sim 30$ K.
The large line widths of the molecular lines, the
high gas phase abundance of  molecules linked to the dust chemistry,
and the difference between
gas and dust temperature suggest that some kind of shocks could be responsible
for the high gas temperatures of the molecular gas  (Wilson et al. 1982,
Martín-Pintado et al. 2001).
The possible influence of radiation in the heating of the molecular gas
is also  usually ruled out due to the lack of far infrared and thermal
radio continuum sources in the GCr others than the well known
H {\sc ii} regions  
associated to the Sgr complexes (A, B,...) or ionized nebulae
like the Sickle.

To investigate the heating of the molecular clouds in the GCr we
have studied a sample of 18 clouds located all along this region.
The clouds were selected as molecular peaks located far from
far infrared or radio continuum sources.
Those  sources were observed with the spectrometers on board the 
{\it Infrared Space Observatory} (ISO) and with the IRAM 30-m telescope.
The data obtained with ISO
are observations of the lowest H$_2$ pure-rotational
lines, dust continuum spectra from 40 to 190 $\mu$m and a number of
fine-structure lines from neutral atoms or ions
(O I 63 $\mu$m, C II 158 $\mu$m, Ne II 12 $\mu$m, O III 52 $\mu$m ...). 
With the  IRAM 30-m antenna we have observed C$^{18}$O, $^{13}$CO, H35$\alpha$
and H41$\alpha$.

In these paper we will review the results already published in
Rodríguez-Fernández et al. (2001a, 2001b, hereafter
RF01a, RF01b, respectively) and we will present
some new results that will be extended elsewhere
(Rodríguez-Fernández et al. 2003, RF03).



\section{Warm molecular gas}        
\label{required}

Before ISO, the warm gas component of the GCr molecular  clouds
was mainly studied by means of NH$_3$ observations.
The multilevel study of Huttemeister et al. (1993) showed that the temperature
structure of the GCr clouds can be characterized  with two gas
components at different temperatures: a cold gas component
with a temperature close to that of the dust ($\sim 20$ K)
and a warm gas component whose temperature ranges from 100 to 250 K.
However, since the abundance of ammonia is known to vary significantly
is difficult to estimate the column density of warm gas in the GCr clouds.
We have, for the first time, obtained the column density of warm 
gas in the GCr clouds by observing H$_2$ pure rotational lines 
(RF01a).
Columns 3 and 4 of  Table 1 show the total column density and the 
temperature of the warm gas, respectively.
We have also estimated the H$_2$ density and 
the total column density of molecular gas in this
clouds by observing $^{13}$CO and C$^{18}$O and doing
a radiative transfer analysis for kinetic temperatures
between 15 and 200 K.
The results are shown in columns 1 and 2 of Table 1.
The fraction of warm H$_2$ to the total H$_2$ column density as traced
by CO varies from source to source but it is of $30 \%$ on average.

As discussed in RF01a, it is difficult
to explain the large column density of warm gas in the GCr  clouds.
Several low velocity ($\sim 10$ Km\,s$^{-1}$) C-shocks,  photo-dissociation
regions (PDRs) or both should be present in the line of sight.
Comparing the energy of the turbulent motions in the GCr clouds with
the cooling by H$_2$ (which at the moderate density of the GCr clouds
is comparable to that by CO), one finds that dissipation of
supersonic turbulence
could account for the heating of the warm H$_2$.
However, with the available data is not possible to rule out heating
in PDRs, indeed the observed temperature gradient in the GCr clouds
can be appropriately reproduced in a context of a PDR (RF01a).



\begin{table}[htb]
{\footnotesize
\caption{Physical parameters derived from the observations. See text
for explanation. Number in parentheses are errors of the last significant
digit. Typical errors of $T_1$ and $T_2$ are 5 and 1 K, respectively.}
\label{tab:2}
\renewcommand{\arraystretch}{1.5}
\begin{center}
\begin{tabular}{llllllllllll}
\hline
Source & $n_{H_2}$ & $N_{H_2}$(CO) & $N_{H_2}$ & $T_{H_2}$ & $L_{FIR}$ &
$T_1$ & $T_2$ & $Q(H)$ & $n_e$ & $T_{eff}$ \\
  & cm$^{-3}$ & cm$^{-2}$ & cm$^{-2}$ & K & W\,cm$^{-2}$&  K &K & s$^{-1}$
  & cm$^{-3}$ & K \\
  &    & $\times 10^{22}$& $\times 10^{22}$ &   & $\times 10^{-15}$ &
  & & &$\times 10$ & $\times 10^3$ \\
\hline
M-0.96+0.13 &3.5-4&0.6-1.1&1.10(9) &157(6) &2.9&30 & 15& $\le$47.8& $\le$2.5 &... \\
M-0.55-0.05 &3.8-4.4&4.3-6.0 &2.7(3)&135(5)&12.3&31 &16& $\le$47.5 &$\le$7.1 & ...\\
M-0.50-0.03 &3.4-3.7&2.4-3.0&2.3(2)&135(4)&10.5& 34 & 17& $\le$48.0 & $\le$2.9 &... \\
M-0.42+0.01 &4-4.5&2.1-3.4&1.03(8)&167(6)&9.8& 30 & 16& $\le$48.0& $\le$30 &...\\
M-0.32-0.19 &$>$3&1.1-2.2&1.03(5)&188(5)&7.7& 35 & 16& $\le$47.6& 4.4 & 34 \\
M-0.15-0.07 &3.7-4.1&6.6-8.4&2.6(4)&136(6)& ...&... & ...& $\le$47.6 & ...  &...           \\
M+0.16-0.10 &3.8-4.2&3.7-4.9&1.17(13)&157(7)&9.9& 32 & 17& $\le$47.3& 9.4 &35 \\
M+0.21-0.12 &$>$3&0.8-1.5&0.64(7)&186(13) &13.5& 39 & 16& $\le$47.6 & 15&35\\
M+0.24+0.02 &3.5-3.9&4.8-7.1&1.73(6)&163(2)&16.1& 30 & 16& $\le$47.4& 2.5 &...\\
M+0.35-0.06 &3.8-4.8&1.7-2.7&0.66(5)&195(11)&14.5& 36 &18& $\le$48.1& $\le$20 &...\\
M+0.48+0.03 &3.8-4.2&3.2-3.6&1.03(9)&174(7)&9.8& 26 &14& $\le$47.5 & $\le$16 & ...\\
M+0.58-0.13 &3.6-3.9&3.1-3.9&1.3(2)&149(5)& 7.2&27 &15& $\le$47.5 &  $\le$2.5 & ...\\
M+0.76-0.05 &2.8-4.2&6.6-8.6&1.77(8)&181(4)&8.1& 24 & 15& $\le$47.4 & 11 & ...\\
M+0.83-0.10 &3-4.5&4.8-6.5&1.59(6)&178(5)&6.3& 26 & 15& $\le$48.0 &  $\le$86 & ...\\
M+0.94-0.36 &$>3$ &1.3-2.9&0.95(10)&146(7)&...& ...&...& $\le$48.0 & ...     &...        \\
M+1.56-0.30 &3-4.5&2-0.5&0.26(10)&260(30)&1.5& 27 & 17& $\le$48.0 & $\le$140 &...\\
M+2.99-0.06 &3-4 &1.0-2.1&1.40(9)&152(3)&1.7& 29& 18& $\le$48.0 & $\le$27 &...\\
M+3.06+0.34 &3-3.8&1.6-1.2&0.21(8)&250(20) &1.1&34& 18& $\le$48.0 & $\le$140 &... \\
\hline
\end{tabular}
\end{center}
}
\end{table}

\section{Dust temperatures}        
\label{required}
\begin{figure}[htb]
\includegraphics[width=6cm]{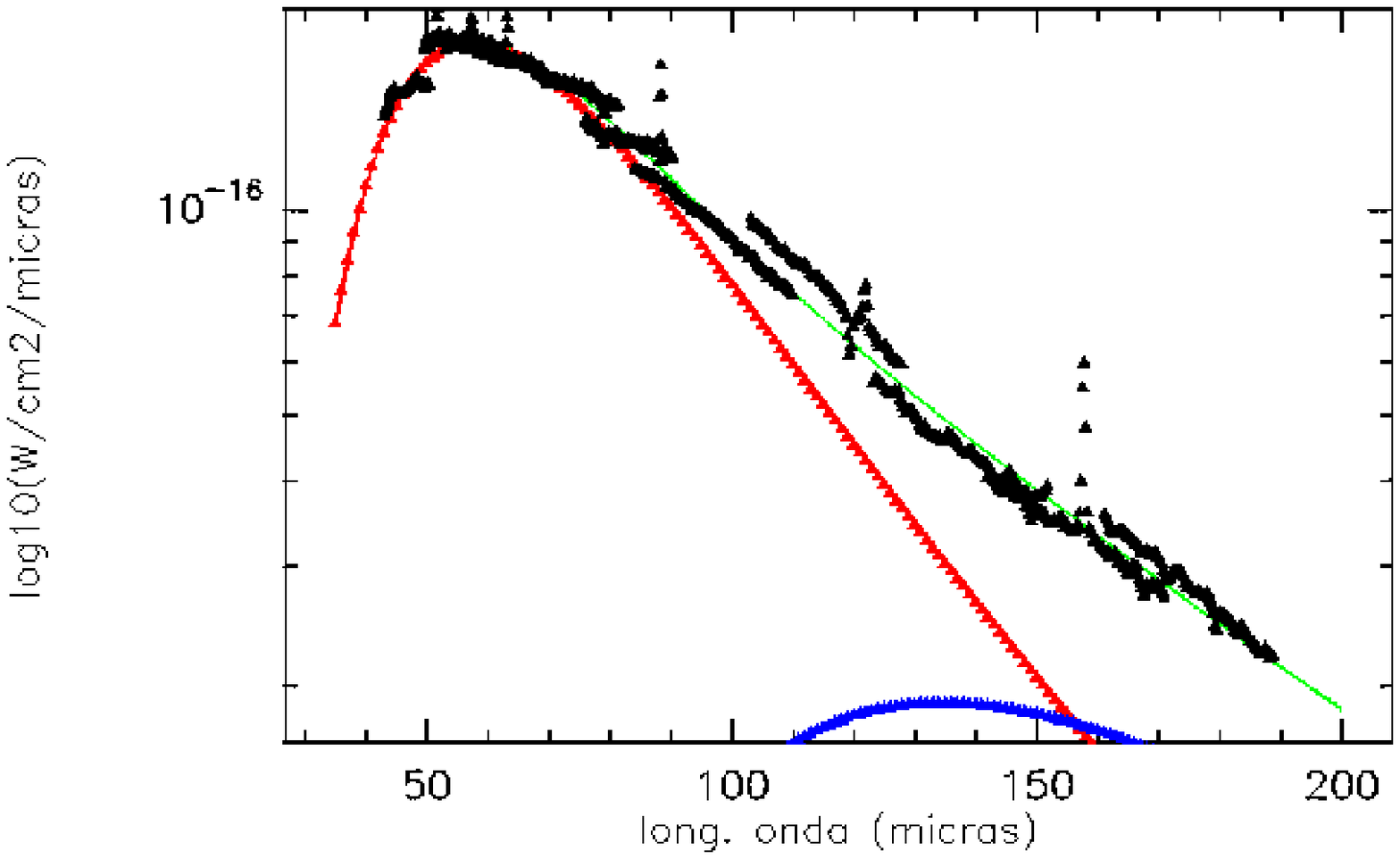}
\hfil
\includegraphics[width=6cm]{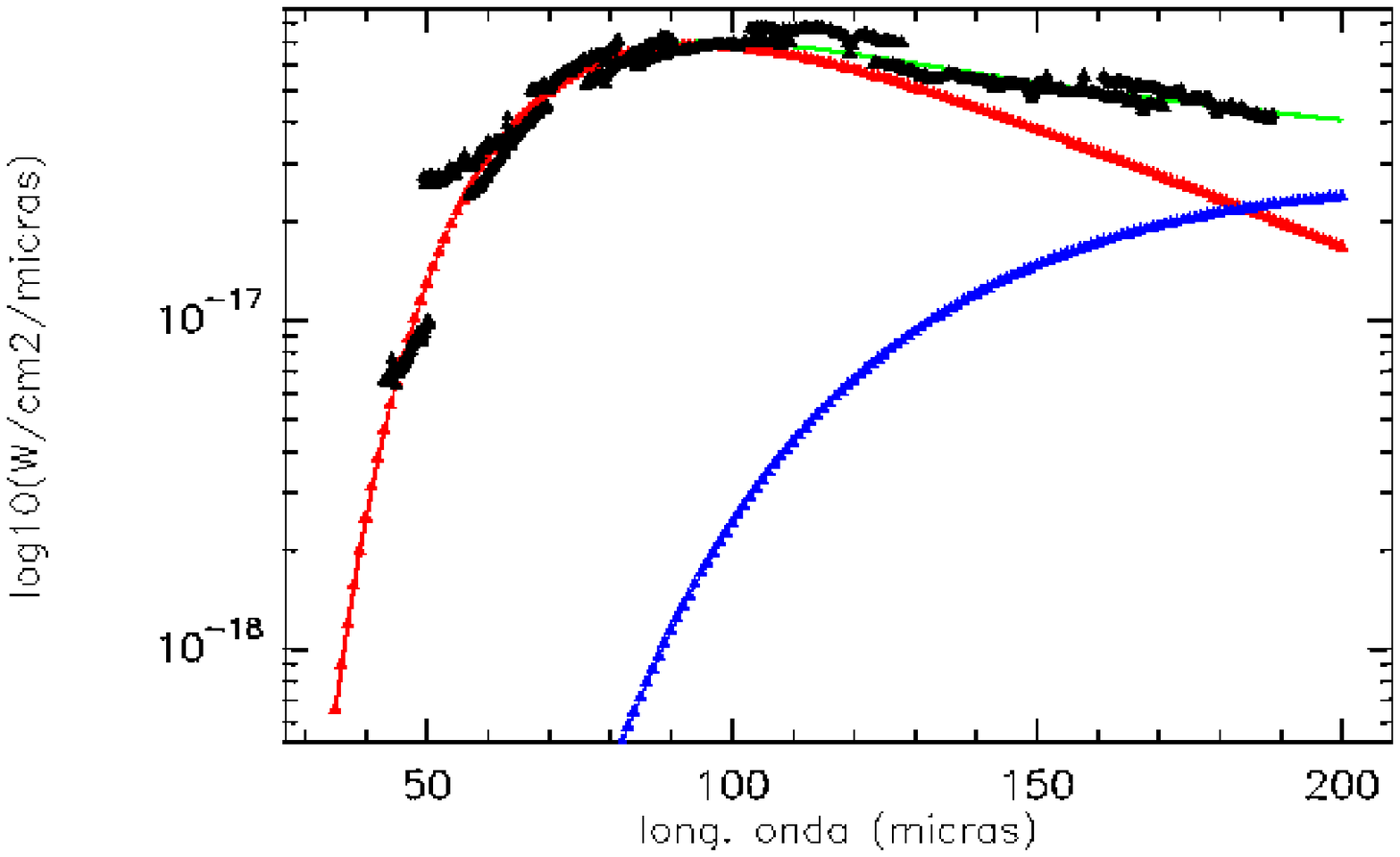}
\caption{Left panel: Dust continuum emission towards M+0.21-0.12 (black
solid triangles) and grey-body fit (green) with two temperature
components ar 39 (red) and 16 K (blue). Right panel: same for M+0.76-0.05
with a 24 K (red) and 15 K (blue) components.}
\label{rodriguez_1_f2}
\end{figure}

The dust continuum  emission peaks at wavelengths of 100 to 80 $\mu$m for all
but two sources, whose spectra peaks at 50-60  $\mu$m. 
The dust luminosity in the observed range is listed in column 5 of Table 1.
It is not possible to fit the spectra with just one grey body.
Thus we have used a model with two grey bodies like that described in
section 2 of Goicoechea et al. (2003).
Figure \ref{rodriguez_1_f2} shows the data and the grey bodies fits for
two sources.
To explain the emission at long wavelengths it is needed a  dust component
with a temperature of 15 to 18 K.
The temperature of the warmer component varies from source to source
from 26 to 39 K.
Due to the uncertainties in the dust emissivity it is not easy
to determine a total column density of dust 
(on the contrary, temperatures are almost independent on the dust emissivity).
Nevertheless, the column density of dust with temperatures higher than
50 K is less than 500 times lower than the column of dust at 15-35 K.


\section{Radio recombination lines}        
\label{required}
To study the possible presence of ionized gas associated with the
GCr molecular clouds we have observed the
H41$\alpha$ and H35$\alpha$ recombination lines using the 
IRAM 30m antenna (RF03).
However, we have not detected any of the lines in any of the sources.
From our 3$\sigma$ limits to the line fluxes we have determined upper
limits to the flux of Lyman continuum photons in the 30m beam (column 8 of
Table 1).
The comparison of those limits with stellar atmospheres models
rules out the presence of stars of an earlier type than B0.5
and effective temperatures higher than $\sim$ 32000 K.


\section{Fine structure lines}        
\label{required}

We have detected lines of atoms and low excitational potential ($<13.6$ eV)
ions like O I 63 $\mu$m, C II 158 $\mu$m or Si II 34 $\mu$m
in all of the sources.
In most of them we have also detected N II 122 $\mu$m or Ne II 12 $\mu$m 
(excitational potential of 14 and 21 eV respectively).
In 11 of the 18 sources we have even detected the O III 88  $\mu$m 
line (the excitational potential of O III is 35 eV).
Column 9 of Table 1 shows the electron densities of the ionized gas
component as derived from the O III 52 to O III 88 $\mu$m lines ratios.
The densities, around 10-100 cm$^{-3}$,
are lower than those found in the Radio Arc region
(RF01b) and in Sgr B2 (Goicoechea et al. 2003).
Column 10 of Table 1 shows the effective temperatures derived from
the N III 57 $\mu$m  to N II 122 $\mu$m lines ratios for the three sources
were the N III line has been detected.
These temperatures have been calculated following the
H II regions models by Rubin et al (1994).
However,
as pointed out by Shields and Ferland (1994)
the observed  lines ratios can even be reproduced
with higher effective temperatures of the ionizing radiation if the
ionization parameter is low, i.e., if the
medium is clumpy and inhomogeneous and the ionizing sources are located
far from the ionized nebulae.
These seems to be the case of the Radio Arc region, were we
have shown the presence of an extended component of 
gas ionized by the combined effect of the Quintuplet and the Arches clusters
(RF01b).
The radiation can reach large distances due to the inhomogeneity of the 
medium (in part due to the presence of a large bubble clearly seen in 
infrared images). 
Fotoionization model calculations 
showed that the lines ratios observed in this region
can be explained with a constant effective temperature of the ionizing
radiation but a different ionization parameter for each cloud
consistently with
the different distances of the clouds to the ionizing sources.

The analysis is more difficult for 
the clouds located far from the Radio Arc region since the geometry of the
medium and the possible ionizing sources are unknown.
However, fotoionization simulations for many lines ratios demonstrate that 
effective temperatures of $\sim 35000$ K are possible if the ionization
parameter is low ($\sim 10^{-3}$, RF03).

\begin{figure}[htb]
\begin{minipage}[t]{.50\textwidth}
\includegraphics[width=0.6\textwidth]{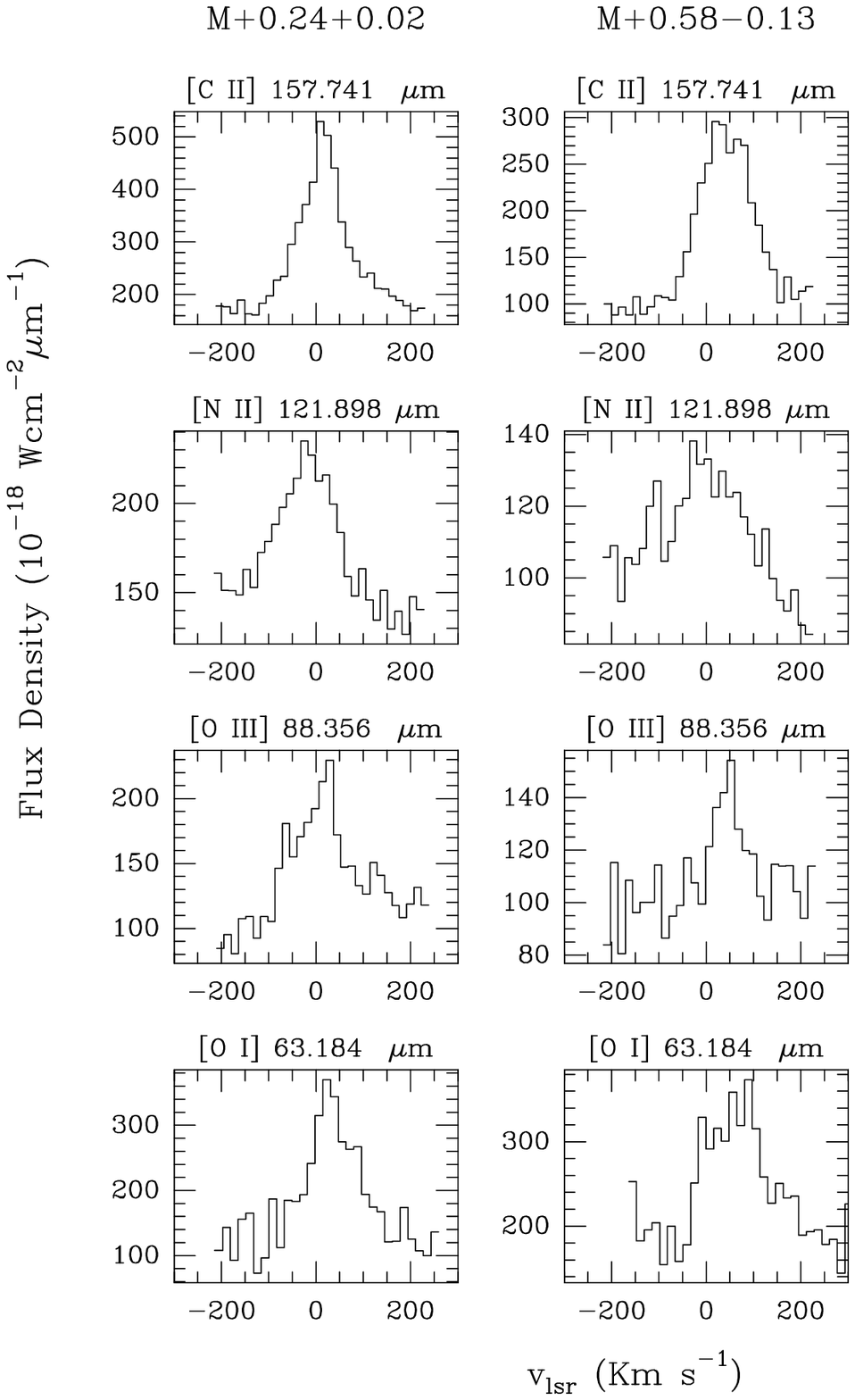}
\caption{Fabry-Perot spectra}
\label{rodriguez_1_f1c}
\end{minipage}
\hfill
\begin{minipage}[t]{.40\textwidth}
\vspace{-6.5cm}
\begin{center}
\includegraphics[width=.5\textwidth]{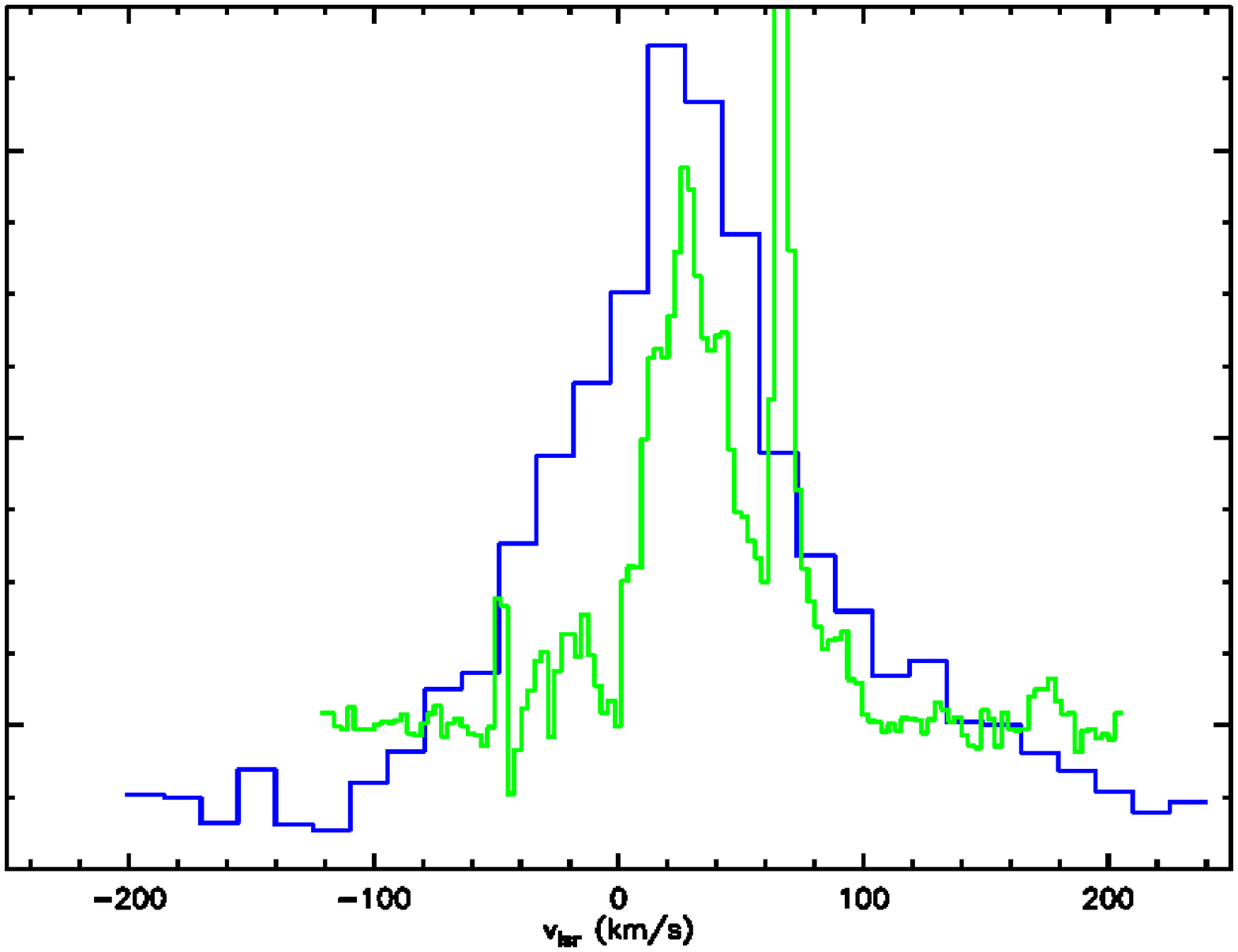}
\includegraphics[width=.5\textwidth]{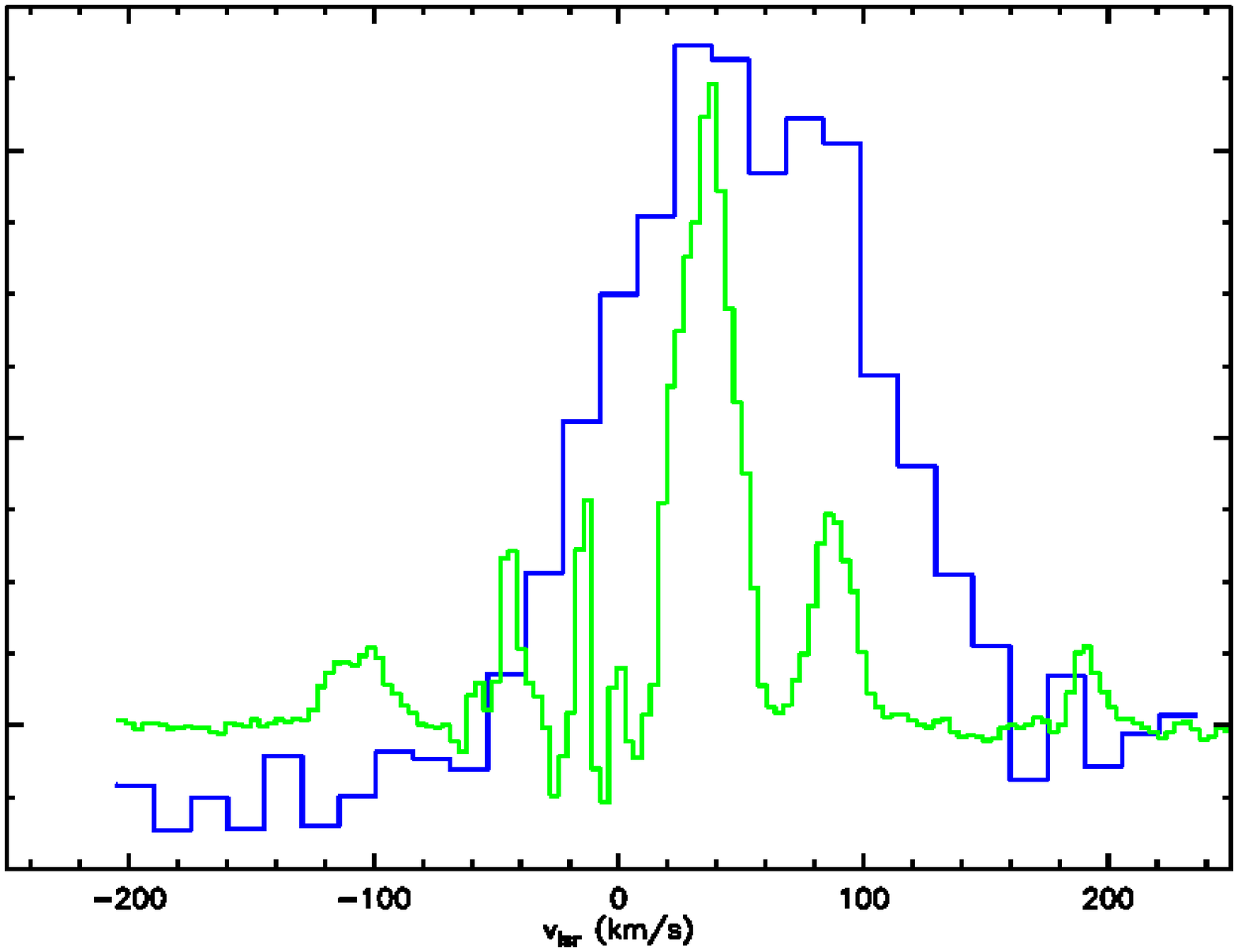}
\end{center}
\caption{A comparison of the line profiles of $^{13}$CO(1-0) (green) and
C II 158 $\mu$m (blue) towards M+0.24+0.02 (up) and M+0.58-0.13 (down).
Intensity in arbitrary units.}
\label{rodriguez_1_f1a}
\end{minipage}
\end{figure}

Some of the fine structure lines have been observed in the Fabry-Perot
mode. The spectral resolution of this mode ($\sim $ 30 km\,s$^{-1}$) 
give us the possibility of studying the line profiles of the broad
GCr lines.
Figure \ref{rodriguez_1_f1c}
shows a sample of the lines observed in this mode towards two sources.
Taking into account the moderate spectral resolution,
the lines profiles and the lines centers of highly excited ions like
O III are in good agreement with those of the neutral or low excitational
potential ions like (O I and C II).
Furthermore, Fig.  \ref{rodriguez_1_f1a}
shows that the agreement of the line profiles
of the weakly ionized gas and the molecular gas is rather good.
This fact suggest that the three components are associated and 
that radiation could play a role not only in the ionization of
the ionized gas but also in the heating of the neutral gas.

We have compared the observed far infrared continuum and the 
C II, O I and Si II 
lines fluxes with the predictions from PDR models. 
The power  radiated by lines is $\sim 0.5 \%$ of that radiated by
the continuum. 
While the C II/O I ratio is $\sim$ 5.
Plotting both quantities in a plot like that of Fig.~2d of 
Goicoechea et al. (2003), one finds that
the GCr clouds exhibit similar properties
to those of Sgr B2, with a far ultra-violet field $\sim$ 10$^{3}$ times
larger  than the local interstellar radiation field and a hydrogen
density of $\sim 10^{3}$ cm$^{-3}$.
The absolute fluxes of the C II  ($\sim 10^{-17}$ W\,cm$^{-2}$) and
the O I line ($\sim 2\,10^{-18}$ W\,cm$^{-2}$)
are also consistent with the Hollenbach et al. (1991) PDRs models
predictions.

\section{Discussion: heating and ionization}
\label{required}
The IRAM-30m ``radio view''
of the GCr interstellar medium seems to confirm
the ``classical'' idea of a mainly neutral and dense  gas.
However, 
the global picture arising from the ISO observations is a complex 
interstellar medium with associated molecular, atomic, and
ionized gas components
with decreasing densities ($n_{H_2}\sim 10^4, n_H \sim 10^3,
n_e \sim 10^{1-2}$ cm$^{-3}$).
The ionizing radiation is hard  (effective temperatures close to 
$\sim 35000$ K) but diluted due to the large 
distances from the ionizing sources to the nebulae ($\sim 50$ pc in
the Radio Arc region).
The large distance effect of the ionizing radiation can only be explained
if the interstellar medium is very inhomogeneous.
This scenario is also necessary to explain the low number of Lyman
continuum photons derived from the radio recombination lines observations.

The radiation must also influence the atomic and molecular phases.
As we have seen, the O~I and C~II lines are well explained by a PDR
with a H density of $\sim 10^3 $ \, cm$^{-3}$  and a far ultra-violet
incident field 10$^3$ times higher than that in the local interstellar
medium.
The absolute O I and C II lines fluxes predicted by the PDR models
are also similar  the observed fluxes arising from the GCr clouds.
Those models also predict a warm H$_2$ layer with temperatures of 
$\sim 150$ K, as we have derived from the pure-rotational lines. 
However, the total column density of warm H$_2$ predicted by the models
is $\sim 3\,10^{21} $ \, cm$^{-2}$ 
while the  total column density of warm H$_2$ that we have derived from the
pure-rotational lines is $\sim 10^{22} $ \, cm$^{-2}$.
Thus, we conclude that approximately 30 $\%$ of the warm molecular gas in the
GCr clouds arises in PDRs in the external layers of the clouds.

It is important to note that
the discrepancy of dust and gas temperatures only rules out gas heating
by gas collisions with hot dust, but it does not rule out {\it all}
radiative heating mechanisms.
In the external layers of the proposed   PDRs the gas is heated to 
temperatures of $\sim 150$ K by photo-electric effect on the dust grains
without heating the dust to temperatures higher than $\sim 35$ K
(Hollenbach et al. 1991).
In fact, the $\sim 30$~K dust component in the GCr can be associated
to the 150 K gas.

At least a fraction of the other 70 $\%$ of warm gas should be heated 
by shocks.
The main evidence for shocks in the GCr are the high degree of turbulence
revealed by the large line-widths and the high abundance in gas phase of 
molecules linked to the dust chemistry as
SiO, NH$_3$ or C$_2$H$_5$OH (Martín-Pintado et al. 2001),
which are easily photo-dissociated in the presence of ultra-violet
radiation.


\begin{acknowledgement}
NJR-F has has been supported by a Marie Curie Fellowship of the European
Community program ``Improving Human Research Potential and the Socio-economic
Knowledge base'' under contract number HPMF-CT-2002-01677.
NJR-F acknowledges useful discussions with J.R. Goicoechea.
\end{acknowledgement}

\end{document}